# ChemComm

## COMMUNICATION

### Why is BeGeN$_2$ different? A computational bonding analysis

Joachim Breternitz,*[a]

**Until recently, all known ordered II-IV-N$_2$ materials crystallised in the β-NaFeO$_2$-type in space group *Pna*2$_1$. BeGeN$_2$, however, crystallises in space group *Pmc*2$_1$, with a different cation ordering motif long anticipated for this materials class. Chemical bonding analysis is used to rationalise this particular behaviour.**

Nitride materials based on the wurtzite type, particularly In, GaN, AlN and their mixtures are known for their electronic and optoelectronic properties.[1–4] Mainly in response to the scarcity of In and Ga, alternative formulations of wurtzite-based nitrides have been investigated, in which the trivalent ions Al$^{3+}$, Ga$^{3+}$ and In$^{3+}$ are replaced by different heterovalent substitutions, leading to a whole series of new stoichiometries.[5–8] The crystal structures of these ternary and multinary compounds are still closely related to the wurtzite-type crystallising in subgroups of the *P*6$_3$*mc* wurtzite-type structure.[9,10] Interestingly, the increased chemical and structural complexity of these materials also comes at the benefit of opening a novel mechanism for tuning the optoelectronic properties through cation disorder.[11]

Probably the most-researched materials contain a divalent and a tetravalent cation in a 1:1 ratio, leading to a general stoichiometry of II-IV-N$_2$. Some of the reported materials are cation-disordered and crystallise in the parent wurtzite-type, but this is not the thermodynamically stable case. Virtually all materials reported so far crystallise in the β-NaFeO$_2$-type (space group *Pna*2$_1$). The unit cell volume of this ordered structure type is 4 times larger than the wurtzite type, allowing for two distinct cation positions. One of the striking advantages of this cation arrangement is the preservation of Pauling's 2$^{nd}$ rule, also dubbed as octet rule (figure 1): Every nitrogen atom is surrounded by two divalent and two tetravalent anions, allowing for local charge compensation.

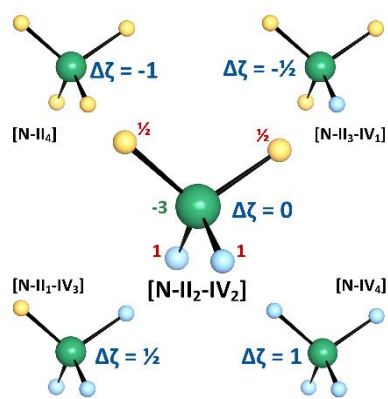

Figure 1: Representation of the different local anion coordination environments in II-IV-N$_2$ nitrides. The strengths of the electrostatic valence bonds for the cations (s, red numbers) and the charge of the anions (z, green number) are given for the Pauling rule preserving case (centre). The sum Δζ = Σs + z is given for all of the possible tetrahedra. Adapted from [12].

Still, there is a second cation ordering motif in space group *Pmc*2$_1$, which also preserves Paulings rule.[13–18] Since its unit cell is only twice as large as the wurtzite one, it is nominally of higher symmetry than the β-NaFeO$_2$-type, which often raised the question, why it was not observed experimentally. Particularly in systems like ZnSnN$_2$, where the calculated energy difference is only 13 meV/f.u.[19] From a structural point-of-view, we argued in an earlier paper that the *Pmc*2$_1$ structure did not form due to the crystallographic restraints,[9] that forced the different tetrahedra† to have equal sizes or deviate strongly from the ideal tetrahedral shape. Very recently, however, Krach et al. reported on the synthesis of BeGeN$_2$, which crystallises in the long-anticipated structure type in space group *Pmc*2$_1$,[20] which should consequently be denoted as BeGeN$_2$-type. Given this surprising experimental finding, it is the intention in this work to understand the underlying reason and differences between BeGeN$_2$ and other materials in order to rationalise, why BeGeN$_2$ is structurally distinct.

*a.* FH Münster University of Applied Sciences, Stegerwaldstraße 39, 48565 Steinfurt, Germany. Joachim.Breternitz@fh-muenster.de

Supplementary Information available: [details of any supplementary information available should be included here]. See DOI: 10.1039/x0xx00000x

Table 1: Conclusion of the main values for the different compounds studied in this work.

| Compound | $\Delta E_{tot}$ /(meV/f.u.) | < (N-$M^{II}$-N) $Pmc2_1$ | < (N-$M^{IV}$-N) $Pmc2_1$ | ICOBI $M^{II}$-N $Pmc2_1$($Pna2_1$) | ICOBI $M^{IV}$-N $Pmc2_1$($Pna2_1$) | ICOHP – [$M^{II}N_4$] /eV $Pmc2_1$ | $Pna2_1$ |
|---|---|---|---|---|---|---|---|
| BeSiN$_2$ | 10 | 108.7°-110.1° | 108.0°-110.4° | 14.9 % (15.1 %) | 79.1 % (78.9 %) | -28.15 | -28.14 |
| BeGeN$_2$ | -7 | 108.7°-114.5° | 106.0°-111.0° | 16.0 % (16.0 %) | 81.6 % (81.4 %) | -25.69 | -25.55 |
| MgSiN$_2$ | 276 | 95.3°-120.0° | 107.1°-113.7° | 14.4 % (14.2 %) | 82.3 % (82.7 %) | -27.40 | -27.76 |
| MgGeN$_2$ | 135 | 101.0°-116.2° | 107.5°-110.5° | 15.4 % (15.1 %) | 84.3 % (84.7 %) | -25.09 | -25.48 |
| ZnGeN$_2$ | 84 | 102.0°-112.7° | 108.8°-110.8° | 50.1 % (49.9 %) | 77.9 % (78.1 %) | -24.38 | -24.68 |
| *MgPbN$_2$* | -0.3 | 108.0°-114.4° | 107.3°-111.4° | 15.9 % (15.9 %) | 75.7 % (75.9 %) | -18.42 | -18.48 |
| *ZnPbN$_2$* | -2 | 106.8°-117.4° | 105.8°-111.4° | 53.4 % (53.5 %) | 67.6 % (67.6 %) | -16.85 | -16.79 |

In a first step, the crystal structures for BeGeN$_2$ in the BeGeN$_2$-type and the β-NaFeO$_2$-type were optimised. The lattice constants match the reported ones[20] within an accuracy of 0.5 %. Furthermore, we performed the same calculations for BeSiN$_2$, MgSiN$_2$, MgGeN$_2$ and ZnGeN$_2$, which are all known to crystallise in the β-NaFeO$_2$-type.[21–25] The total energies of the different calculations confirm that the BeGeN$_2$-type is thermodynamically stable for that compound with an energy difference of 7 meV/f.u. compared to the β-NaFeO$_2$-type. This is in contrast to all other tested compounds, where the β-NaFeO$_2$-type is thermodynamically stable with larger margins (table 1). This confirms the solitary position of BeGeN$_2$ in the adoption of the BeGeN$_2$ type. It is worth mentioning that the optimised volumes per formula unit show a similar trend in that the one for BeGeN$_2$ is slightly larger in the BeGeN$_2$-type (Diff. 0.03 Å³), while the trend for all other compounds is inverse with the β-NaFeO$_2$-type having the larger volumes per formula unit. (V($Pmc2_1$)-V($Pna2_1$): BeSiN$_2$: -0.12 Å³/f.u.; BeGeN$_2$: 0.03 Å³/f.u.; MgSiN$_2$: -0.12 Å³/f.u.; MgGeN$_2$: -0.19 Å³/f.u.; ZnGeN$_2$: -0.12 Å³/f.u.).

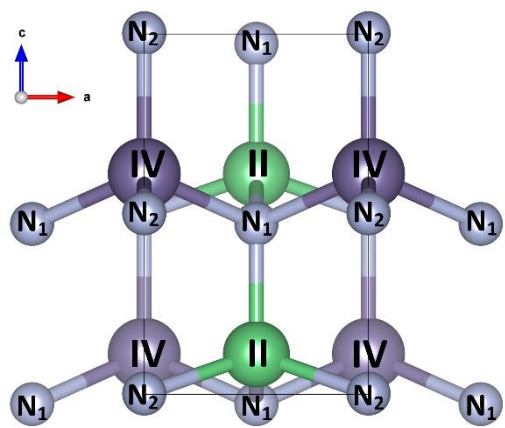

Figure 2: Coordination environment in the BeGeN$_2$-type depicted on the optimised crystal structure of BeGeN$_2$.

This alone, however, is not sufficient to explain the discrepancy between BeGeN$_2$ and the other compounds and a closer look into the coordination environment within the BeGeN$_2$-type (figure 2) is appropriate.

In an earlier publication, we argued that from a geometrical point-of-view, the BeGeN$_2$-type should be disfavoured, since the mirror planes in space group $Pmc2_1$ at $a$ = 0 and $a$ = ½ enforce restrictions on the bond lengths and angles. An undistorted network of tetrahedra would, therefore, require the respective cations to have equal size. Regarding the bond lengths and angles, this is indeed the case, as BeSiN$_2$ appears to exhibit the least distorted bonding angles (table 1). Nonetheless, this compound crystallises in the β-NaFeO$_2$-type, while BeGeN$_2$ crystallising in the BeGeN$_2$-type appears slightly more distorted (table 1). Taking the other compounds into account, as well, it becomes more evident that the divalent cation position becomes more and more distorted, the larger the difference between divalent and tetravalent cation gets. This peaks for MgSiN$_2$ (Shannon radii: Mg$^{2+}$: 0.57 Å, Si$^{4+}$: 0.26 Å)[26] with N-Mg-N angles between 95.3° and 120.0°. It is remarkable that the tetravalent cation tetrahedra are much less distorted, indicating that they are probably much more structure determining (table 1).

Krach et al. suggest that the BeGeN$_2$-type is essentially formed, when the divalent cation is smaller than the tetravalent cation.[20] To verify this point, one can perform a thought experiment in comparing the total energies of analogue compounds with larger tetravalent cations. In this case, we calculated the total energies for *MgPbN$_2$* and *ZnPbN$_2$* in both structure types. It needs to be highlighted that this is a thought-experiment and does neither suggest that these compounds exist, nor that they would crystallise in any of the two structure types in question, should they exist. In fact, a very recent computational study suggests that both compounds should rather crystallise in the chalcopyrite-type.[27] Nonetheless, the comparison of β-NaFeO$_2$-type and BeGeN$_2$-type structures suggest that the latter is indeed more stable (Energy differences 0.3 meV/f.u. for *MgPbN$_2$* and 2 meV/f.u. for *ZnPbN$_2$*), underlining the argument by Krach et al.

**Journal Name** COMMUNICATION

It remains the question, why not only the existence of a size difference between the two cation types directs the structure formation, but also the question, whether the divalent or the tetravalent cation is smaller. It is, therefore, useful to regard the nature of the respective *M*-N bonds, for which one can employ the integrated crystal orbital bond index (ICOBI) translating to an index for the covalency of a chemical bond (table 1).[28] Interestingly, the tetravalent cations all show a large fraction of covalent bonding, while the divalent cations are predominantly ionically bound to nitrogen – with the exception of $Zn^{2+}$, where the ICOBI is ≈ 50 %. Therefore, it may be more appropriate to think of these materials as nitridosilicates and nitridogermanates, reflecting the differences in bonding nature between the different cation types.

Given the different nature in ionic vs. covalent bonding – undirected vs. directed bonding interaction – it is straightforward to rationalise, why the [$M^{IV}N_4$] tetrahedra appear less distorted than the [$M^{II}N_4$] tetrahedra, which again underlines the structure-directing property of the [$M^{IV}N_4$] tetrahedra and explains, why it matters for the formation of the $BeGeN_2$-type, whether the divalent or the tetravalent cations are smaller. If the divalent cations are smaller than the tetravalent ones, the [$M^{IV}N_4$] can remain largely undistorted, keeping the angle sensitive covalent bonds close to their optimal values. The [$M^{II}N_4$] tetrahedra, on the other hand, are distorted but the energy penalty for this is much smaller, due to the undirected ionic bond between N and $M^{II}$. Therefore, this scenario can be favourable against the lower symmetry β-$NaFeO_2$-type.

Finally, this is supported by regarding the bonding energies as indicated by the integrated crystal orbital Hamilton population (ICOHP).[29,30] It appears here that the [$M^{IV}N_4$] bonding energy is the decisive factor that stabilises the $BeGeN_2$-type for that compound. This trend is consistent for all compounds in this investigation with two notable exceptions: 1) The [$SiN_4$] ICOHP in $BeSiN_2$ is virtually identical for the $BeGeN_2$-type and the β-$NaFeO_2$-type, which is mainly due to the very similar ionic radii of $Be^{2+}$ and $Si^{4+}$,[26] rendering the tetrahedra largely undistorted in both crystal structures. 2) The [$PbN_4$] ICOHP in $MgPbN_2$ is slightly less favourable in the $BeGeN_2$-type than in the β-$NaFeO_2$-type, which may be an artefact given that both structures are not thermodynamically stable and have an energy difference that is almost negligible.

Interestingly, one of the three different Be-N bonds in $BeGeN_2$ (the Be-$N_2$ bonds are equal through the space group symmetry, see fig. 2) contributes antibonding states at the valence band maximum (fig. 3), a feature that is also observed in $ZnGeN_2$.[31] In contrary to this, the Ge-N bonds are all bonding throughout the energy range and their relative contribution is energetically much more significant than the Be-N bonds. This is a common motif within the different models and compounds, and also found in the β-$NaFeO_2$-type structures in a similar manner. In fact, the projected COHP curves for the $BeGeN_2$-type and the β-$NaFeO_2$-type are very similar and do not provide a striking feature that favours one over the other (fig. 4)

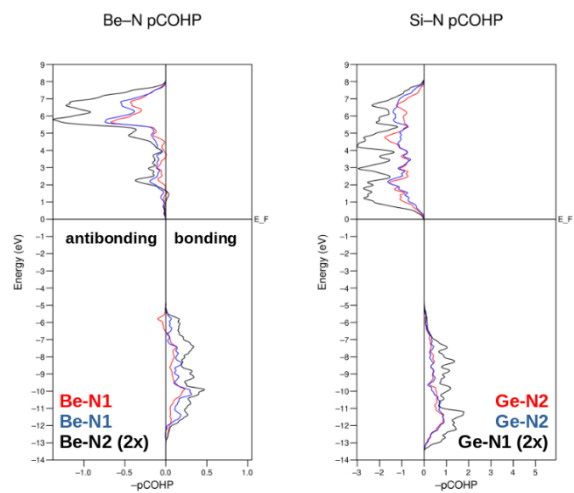

Figure 3: Projected COHPs for the Be-N (left) and Ge-N (right) bonds within $BeGeN_2$ in the $BeGeN_2$-type.

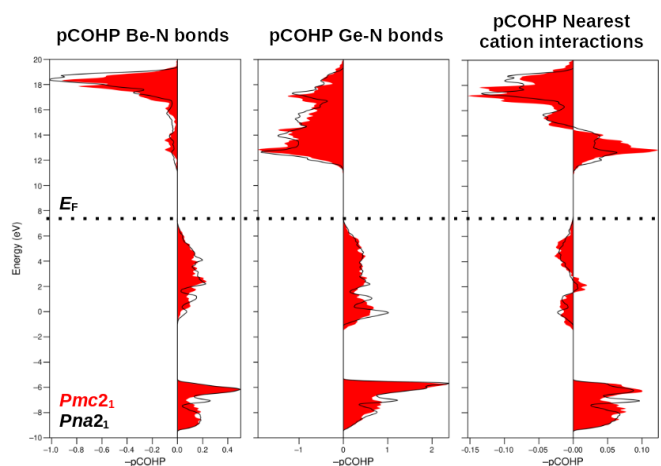

Figure 4: Projected COHPs comparing the averages of all Be-N bonds (left), Ge-N bonds (middle) and nearest cation-cation interactions (right) between the BeGeN2-type (red) and β-NaFeO2-type (black).

Since the cation ordering is the signifying feature that distinguishes the two structure types, it is worth to compare the interactions between neighbouring cations, in order to evaluate whether these may tip the energy balance into one or the other structural type (fig. 4, right). Remarkably, the curves are very similar and although they bear important antibonding fractions towards the top of the valence band, all cation-cation interactions in all tested compounds are overall energetically favourable with negative ICOHPs – some of which are as significant as 25 % of the $M^{II}$-N ICOHPs.

In conclusion, it is apparent that the strength of the $M^{IV}$-N interaction, which is predominantly covalent, is the directing feature for one or the other structure type. In the case of $BeGeN_2$, this is stabilising the crystal structure in space group $Pmc2_1$, which has not been observed before and should be consequently denoted as $BeGeN_2$-type. This compound is probably unique in this structure type.

The calculations in this work were performed with quantum espresso[32–35] using PBEsol exchange-correlation pseudopotentials.[36] Bonding analyses were performed using LOBSTER.[37,38] Further computational details may be found in the SI.

## Conflicts of interest

There are no conflicts to declare

## Data availability

Data for this article including the computational input files and major output files are available at zenodo at http://dx.doi.org/10.5281/zenodo.13998655

## Notes and references

The author would like to acknowledge the University of Münster, where calculations for this publication were performed on the HPC cluster PALMA II, subsidised by the DFG (INST 211/667-1).

† It should be noted that the term tetrahedron is not strictly correct here, since the polyhedron is not tetrahedral through the crystallographic symmetry. Instead, we use the term to describe a sphenoid of approximately tetrahedral shape with this term.

Supplementary Information for

# Why is BeGeN$_2$ different? A computational bonding analysis

Joachim Breternitz

1) Computational details

The crystal structures of all compounds within the study were first geometrically optimised using the BFGS algorithm implemented in Quantum Espresso. Structures were optimised using the respective space group symmetry, and with convergence criteria on energy (< 10$^{-6}$ Ry), force (< 10$^{-4}$ Ry/Bohr) and cell pressure (< 10$^{-2}$ kbar).
SCF calculations as LOBSTER input were then prepared using time reversal symmetry only.
All calculations were uniformly performed using PBEsol exchange-correlation potential PAW pseudopotentials with uniform kinetic energy cutoff (70 Ry) and charge density cutoff (700 Ry). All calculations were performed using a 12 x 12 x 12 Monkhorst-Pack k-point grid.

LOBSTER calculations were performed in a range of -35 eV to +15 eV in order to sample all valence electrons within the partition. Fits were done to the pbeVaspFit2015 basis set using the recommended basis functions as automatically selected by Lobster. *M-M* bonds up to 4 Å were considered for the calculation of the next nearest neighbour interactions.

2) COHP Plots
The colour choices are the same as in the main text figure 3

BeSiN$_2$ – *Pmc*2$_1$

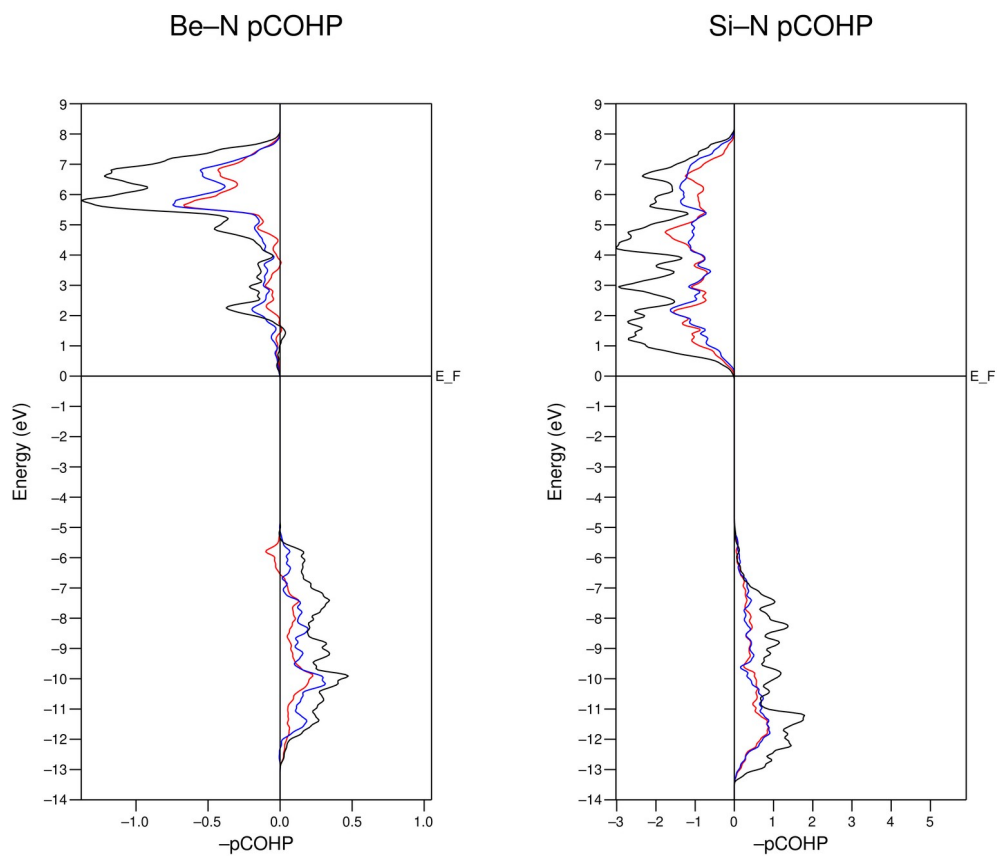

MgGeN$_2$ – *Pmc*2$_1$

## Mg–N pCOHP

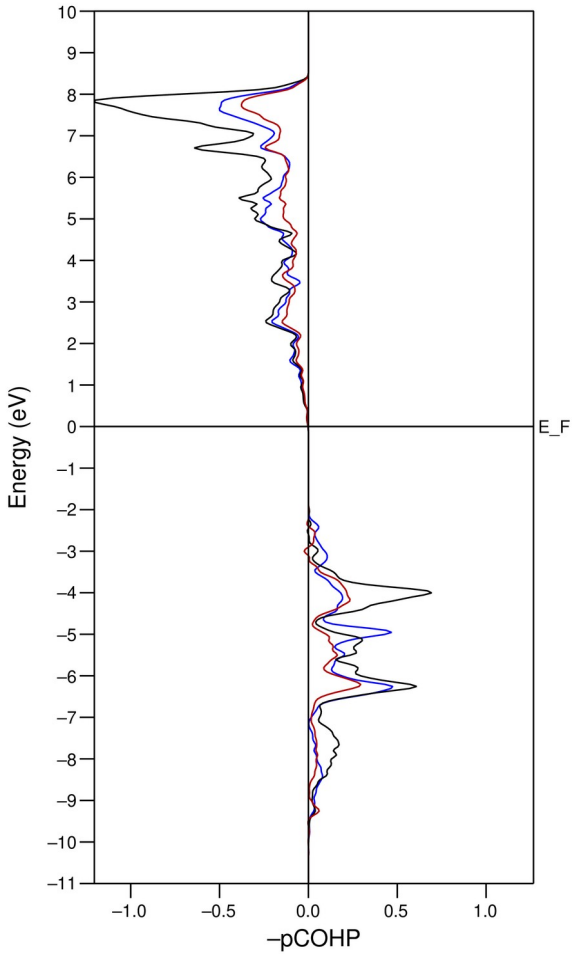

## Ge–N pCOHP

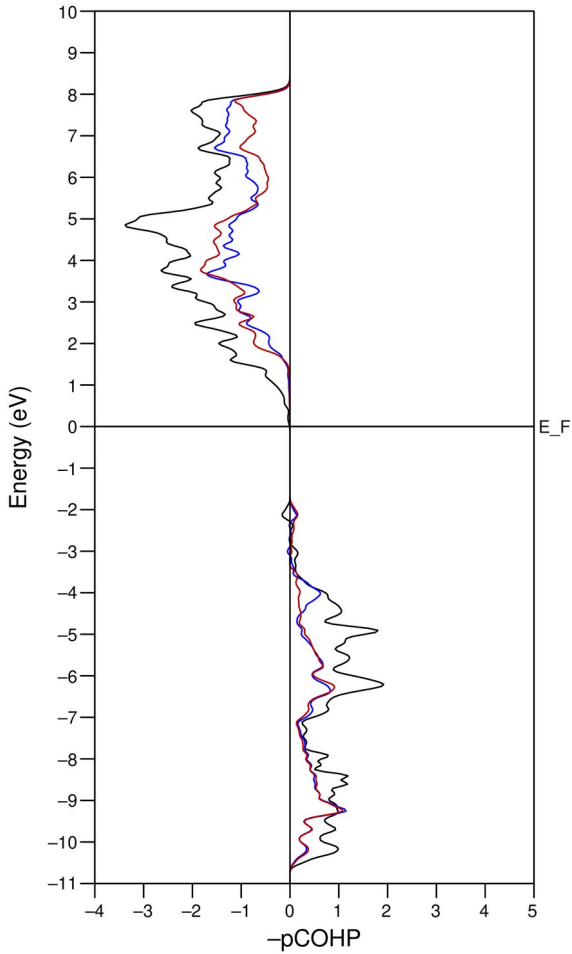